\documentclass[conference]{IEEEtran}

\usepackage{enumitem}
\usepackage[table,dvipsnames]{xcolor}
\usepackage{tcolorbox}
\usepackage{tikz}
\usepackage{amsmath} 
\usepackage{hyperref}
\usepackage{amssymb}
\usepackage{booktabs}
\usepackage{colortbl}
\usepackage{pgf}
\usepackage{etoolbox}
\usepackage{tabularx}
\usepackage{xfp}
\newcommand{\colorize}[1]{%
  \ifdim #1pt<50pt
    \cellcolor{red!\fpeval{100-2*(#1)}!yellow!50}{#1\%}%
  \else
    \cellcolor{yellow!\fpeval{200-2*(#1)}!green!50}{#1\%}%
  \fi
}

\newcommand{\colorizek}[1]{%
  \ifdim #1pt<0.5pt
    \cellcolor{red!\fpeval{100-200*(#1)}!yellow!50}{#1}%
  \else
    \cellcolor{yellow!\fpeval{200-200*(#1)}!green!50}{#1}%
  \fi
}

\newcommand{\step}[1]{%
  \tikz[baseline=(char.base)]{
    \node[shape=circle, fill=black, text=white, inner sep=1pt] (char) {\textbf{#1}};
  }%
}

\newcommand{\joy}{\textit{\textcolor[HTML]{e68a00}{Joy }}} % Dark Yellow
\newcommand{\sadness}{\textit{\textcolor[HTML]{3c78d8}{Sadness }}} % Dark Blue
\newcommand{\anger}{\textit{\textcolor[HTML]{b22222}{Anger }}} % Dark Red
\newcommand{\fear}{\textit{\textcolor[HTML]{76448a}{Fear }}} % Dark Violet
\newcommand{\disgust}{\textit{\textcolor[HTML]{38761d}{Disgust }}} % Dark Green
\newcommand{\surprise}{\textit{\textcolor[HTML]{ff6600}{Surprise }}} % Dark Orange
\newcommand{\anticipation}{\textit{\textcolor[HTML]{993366}{Anticipation }}} % Dark Purple
\newcommand{\trust}{\textit{\textcolor[HTML]{285454}{Trust }}} % Dark Teal
\newcommand{\neutral}{\textit{\colorbox[HTML]{3a3a3a}{\textcolor{white}{Neutral}}}}
\newcommand{\reject}{\textit{\colorbox[HTML]{990000}{\textcolor{white}{Reject}}}}

\newcommand{\joyA}{\textit{\textcolor[HTML]{e68a00}{Joy}}} % Dark Yellow
\newcommand{\sadnessA}{\textit{\textcolor[HTML]{3c78d8}{Sadness}}} % Dark Blue
\newcommand{\angerA}{\textit{\textcolor[HTML]{b22222}{Anger}}} % Dark Red
\newcommand{\fearA}{\textit{\textcolor[HTML]{76448a}{Fear}}} % Dark Violet
\newcommand{\disgustA}{\textit{\textcolor[HTML]{38761d}{Disgust}}} % Dark Green
\newcommand{\surpriseA}{\textit{\textcolor[HTML]{ff6600}{Surprise}}} % Dark Orange
\newcommand{\anticipationA}{\textit{\textcolor[HTML]{993366}{Anticipation}}} % Dark Purple
\newcommand{\trustA}{\textit{\textcolor[HTML]{285454}{Trust}}} % Dark Teal

\newtcolorbox{reviewbox}[1]{%
  boxrule=0.2pt,
  colframe=cyan,
  colback=cyan!3,
  left=1mm,
  right=1mm,
  top=0.8mm,
  bottom=0.8mm,
  boxsep=0pt,
  enlarge left by=0mm,
  enlarge right by=0mm,
  before skip=3pt,
  after skip=3pt,
  width=\linewidth,
  fontupper=\itshape,
  sharp corners,
}

\usepackage{listings}
\begin{document}

%\copyrightyear{2025}
%\copyrightclause{Copyright for this paper by its authors.
%  Use permitted under Creative Commons License Attribution 4.0
%  International (CC BY 4.0).}

%\conference{Joint Proceedings of REFSQ-2025 Workshops, Doctoral Symposium, Posters \& Tools Track, and Education and Training Track. Co-located with REFSQ 2025. Barcelona, Spain, April 7, 2025.}

\title{What About Emotions? Guiding Fine-Grained Emotion Extraction from Mobile App Reviews}
%\title{Guiding the Annotation of Mobile App Reviews for Fine-Grained Emotion Analysis}
%\title{Fine-Grained Emotions in Mobile App Reviews: A Guide to Annotation}
%\title{Emotional Insights from User Feedback: Annotating Mobile App Reviews for Fine-Grained Emotions}

% \author{
% \IEEEauthorblockN{
% Quim Motger\IEEEauthorrefmark{1},
% Max Tiessler\IEEEauthorrefmark{1},
% Marc Oriol\IEEEauthorrefmark{1},
% Xavier Franch\IEEEauthorrefmark{1} and
% Jordi Marco\IEEEauthorrefmark{2}
% \IEEEauthorblockA{\IEEEauthorrefmark{1}\textit{dept. of Service and Information System Engineering (ESSI)}\\ Universitat Polit\`ecnica de Catalunya, Barcelona, Spain\\
% Email: \{joaquim.motger, max.tiessler, marc.oriol, xavier.franch\}@upc.edu}
% \IEEEauthorblockA{\IEEEauthorrefmark{1}\textit{dept. of Computer Science}\\
% Universitat Polit\`ecnica de Catalunya, Barcelona, Spain\\
% Email: jordi.marco@upc.edu}}}

\author{
\IEEEauthorblockN{
Quim Motger, Marc Oriol, Max Tiessler, Xavier Franch, and Jordi Marco}
\IEEEauthorblockA{\textit{Dept. of Service and Information System Engineering (ESSI)}\\ 
Universitat Polit\`ecnica de Catalunya, Barcelona, Spain\\
Email: \{joaquim.motger, marc.oriol, max.tiessler, xavier.franch\}@upc.edu}
\IEEEauthorblockA{\textit{Dept. of Computer Science}\\
Universitat Polit\`ecnica de Catalunya, Barcelona, Spain\\
Email: jordi.marco@upc.edu}
}

\maketitle

\begin{abstract}
Opinion mining plays a vital role in analysing user feedback and extracting insights from textual data. While most research focuses on sentiment polarity (e.g., positive, negative, neutral), fine-grained emotion classification in app reviews remains underexplored. Fine-grained emotion classification is thus needed to better understand users’ affective responses and support downstream tasks such as feature-emotion analysis, user-oriented release planning, and issue triaging. This paper addresses this gap by identifying and addressing the challenges and limitations in fine-grained emotion analysis in the context of app reviews. Our study adapts Plutchik’s emotion taxonomy to app reviews by developing a structured annotation framework and dataset. Through an iterative human annotation process, we define clear annotation guidelines and document key challenges in emotion classification. Additionally, we evaluate the feasibility of automating emotion annotation using large language models, assessing their cost-effectiveness and agreement with human-labelled data. Our findings reveal that while large language models significantly reduce manual effort and maintain substantial agreement with human annotators, full automation remains challenging due to the complexity of emotional interpretation. This work contributes to opinion mining in requirements engineering by providing structured guidelines, an annotated dataset, and insights for developing automated pipelines to capture the complexity of emotions in app reviews.
\end{abstract}

\begin{IEEEkeywords}
  mobile apps,
  app reviews, 
  emotions, 
  opinion mining,
  annotation, 
  dataset, 
  large language models
  %joy \sep
  %trust \sep
  %anger \sep
  %sadness \sep
  %disgust \sep
  %surprise \sep
  %anticipation \sep
  %fear \sep
  %neutral
\end{IEEEkeywords}

\maketitle

\section{Introduction}
\label{sec:introduction}

Mining app reviews from mobile app repositories has gained significant attention in requirements engineering (RE) research over the past decade~\cite{Dabrowski2022}. Relevant descriptors from mobile app reviews include numerical rating~\cite{Dabrowski2023}, review type~\cite{Maalej2015} (e.g., \textit{bug report}, \textit{feature request}, \textit{praise}), topic~\cite{Tushev2022} (e.g., \textit{usability}, \textit{design}, \textit{security}), and polarity~\cite{Dragoni2019} (e.g., \textit{positive}, \textit{neutral}, \textit{negative}). The combined use of these descriptors has led to advanced methods for requirements elicitation~\cite{Pagano2013} and validation~\cite{Al-Subaihin2021} tasks, such as aspect-based sentiment analysis~\cite{Alturaief2021} and feature-based opinion mining~\cite{Guzman2014}.

Among these, polarity has emerged as one of the most widely used descriptors in app review analysis~\cite{Guzman2014,Maalej2015,Bakiu2017,Dabrowski2019,Guzman2019}, and it continues to attract significant attention in recent studies~\cite{Yu2023,Alturayeif2023,Maroof2024}. Polarity is defined as the overall sentiment expressed in user feedback, leading to the classification of textual content into predefined sentiment categories, typically \textit{positive}, \textit{neutral}, or \textit{negative}. %~\cite{Lin2022}. 
Despite its popularity, automatic polarity measurement still presents significant cognitive challenges, such as subtle sentiments, sarcasm, and domain-specific language. These lead to limited precision and low recall in negative feedback~\cite{Dabrowski2023,Alturaief2021}. %While polarity analysis has undergone intense research in software development activities~\cite{Lin2022}, several challenges still remain. 

In addition to these challenges, polarity-based opinion mining lacks the granularity to capture nuanced emotions in user feedback. Polarity labels fail to convey the depth of emotions tied to feature-based opinions, limiting their usefulness for fine-grained analysis. %polarity-based opinion mining is inherently limited in its capacity to provide a fine-grained analysis of the detailed opinions expressed by users. These approaches often fail to capture the subtle nuances and depth of specific emotions conveyed within feedback. As a result, descriptors such as \textit{positive}, \textit{negative}, or \textit{neutral} fall short of fully representing the rich information embedded in user reviews, particularly when analyzing emotions tied to feature-based opinions.
For instance, consider the following \textit{positive} reviews:%\footnote{Examples collected from real user reviews.}:

\begin{reviewbox}
\\[R\textsubscript{1}] Useful app which follows Material Design.
\end{reviewbox}

\begin{reviewbox}
\\[R\textsubscript{2}] Awesome team work but this application does need updating now
\end{reviewbox}

\begin{reviewbox}
\\[R\textsubscript{3}] I run it on my Dropbox cloud storage with Android, Mac and Linux, and I have had no issues
\end{reviewbox}
\vskip 3pt

In addition to its inherent positivity, [R\textsubscript{1}] highlights the user's excitement about a specific characteristic. In contrast, [R\textsubscript{2}] conveys a user request or suggestion for a change or update of the app. Finally, [R\textsubscript{3}] reflects the user's acceptance and personal experience with a particular feature.

Likewise, consider the following \textit{negative} reviews:

\begin{reviewbox}
\\[R\textsubscript{4}] My only complaint is that I sometimes have sync issues with shared notebooks.
\end{reviewbox}

\begin{reviewbox}
\\[R\textsubscript{5}] I really didn't want to make this app my default SMS messaging app on my phone.
\end{reviewbox}

\begin{reviewbox}
\\[R\textsubscript{6}] Very invasive, [...] was forcing to update from a third party store and it wanted to access everything on your phone including your SIM card data.
\end{reviewbox}
\vskip 3pt

In addition to its inherent negativity, [R\textsubscript{4}] conveys minor user disappointment due to issues or bugs with a specific feature. In contrast, [R\textsubscript{5}] reflects the user's outright rejection and decision to stop using the app, which the user considers not suited for purpose. Finally, [R\textsubscript{6}] highlights the user's lack of trust stemming from critical safety and privacy concerns.
%To overcome these limitations, in this research, we explore the hypothesis that a \textbf{more fine-grained, multi-class emotion analysis of app reviews can support software developers and requirements engineers in improving the elicitation and validation of app features through user feedback analysis}. 

This limitation in expressiveness motivates our research. We propose to use \textit{emotion labels} to better capture the nuanced emotional states conveyed in user feedback, allowing more informed and targeted decision-making. This enables use cases such as fine-grained feature-oriented emotion analysis to better prioritize or refine specific app functionalities, and market analysis to categorize user feedback more effectively by emotional trends in terms of user satisfaction and adoption. 

Fine-grained emotion analysis has been explored in opinion mining tasks in various review types, such as products~\cite{Gao2018}, movies~\cite{Sankar2020} and  books~\cite{Lutan2023}. 
However, emotion analysis research in app reviews remains scarce, particularly in the context of mobile app reviews. This entails several limitations in the state of the art: lack of guidelines for human annotators, ignorance on the challenges they face during the annotation process, and lack of public datasets for multiclass emotion extraction, among others. Moreover, the opportunities large language models (LLMs) offer to perform this task remain unknown.  

% highlights the need for curated, open-source datasets to support software engineering tasks~\cite{Hou2024}. To the best of our knowledge, there are no datasets to support fine-grained multiclass emotion extraction from mobile app reviews in the English language.

In this research, we address all these aspects and as a result, we present the following contributions: %\footnote{See \textit{Data Availability Statement} at the end of this paper. % for access to materials, source code and datasets.}:

\begin{enumerate}[label=\textbf{C\textsubscript{\arabic*.}}]
    \item The adaptation of an 8-emotion taxonomy, already used in software engineering, to the context of app reviews.
    \item A set of guidelines and instructions to support the manual annotation of emotions in the context of app reviews. 
    \item A dataset of 1,112 sentences from app reviews annotated with human emotions, belonging to 257 mobile apps.
    \item A set of challenges and design suggestions for the development of automated emotion extraction methods.
    \item Insights into the cost-effectivenes, agreement and correctness of LLM-based annotations with respect to humans.
\end{enumerate}

A replication package including datasets and source code is openly shared (See \textit{Data Availability Statement}). % to ensure transparency and reproducibility. 

Our work lays the groundwork for fine-grained emotion extraction from mobile app reviews, providing a structured dataset, annotation guidelines, and design recommendations for automated methods. By leveraging human annotations alongside LLMs, we assess the feasibility of reducing manual effort in emotion classification while maintaining annotation quality. We expect our findings to guide future research on automating emotion extraction in software reviews more broadly, facilitating its integration into RE processes for improved user feedback analysis.% and decision-making.
\section{Background}
\label{sec:background}

\subsection{Emotion Analysis}
\label{sec:background-emotion}

Recent advances in natural language processing have sparked growing research interest in opinion mining to support software engineering processes~\cite{Lin2022}. Moreover, emotional aspects have been extensively explored in the context of RE tasks~\cite{Hidellaarachchi2023}, including elicitation~\cite{Cheng2023,Taveter2019,Hossein2018}, specification~\cite{Sutcliffe2022} and validation~\cite{Ferrari2024}. Specifically, analysing user feedback from software-related reviews has also emerged as a key approach to integrating emotional aspects into RE~\cite{Araujo2022, Sankar2020,Lutan2023}. These studies vary in scope and purpose, ranging from %investigating human-computer interaction~\cite{Gao2018} to 
analysing app usage experience~\cite{Singh2022} to
classifying emotional states from user feedback on social media platforms~\cite{Khan2024,9582287}. %However, in the context of app reviews, opinion mining research has focused mainly on polarity-based analysis, with less attention given to fine-grained emotions~\cite{Hua2024}. 

Studies on emotion analysis rely on well-established \textit{emotion taxonomies} -- structured classifications of emotions based on psychological theories. These taxonomies vary in scope, granularity, and theoretical foundations across disciplines, making their selection crucial to capturing the necessary level of detail for a given task. While some frameworks define a small set of basic emotions~\cite{Ekman1992, Plutchik2001, Parrott2001}, others introduce finer distinctions with broader emotional states~\cite{Liew2016a}, reflecting diverse perspectives on how emotions are structured and expressed. %The choice of an emotion taxonomy depends on the specific application and dataset characteristics. 
In the context of app reviews, opinion mining has mainly focused on polarity-based analysis~\cite{Guzman2014, Hua2024}, while fine-grained classifications %, which are essential for capturing subtle variations in affective states, 
are more common in other domains such as clinical and psychological studies~\cite{Cowen2019}. Bridging this gap requires adopting emotion taxonomies that balance granularity and applicability to the analysis of app reviews, as explored by prior work (see Section~\ref{sec:related-work}).
%For example, models used for analyzing user feedback in app reviews often rely on coarse-grained polarity-based taxonomies (e.g., \textit{positive}, \textit{negative}, and \textit{neutral})~\cite{Guzman2014}, whereas clinical and psychological studies may require fine-grained classifications to capture subtle variations in affective states~\cite{Cowen2019}.

\subsection{Annotation Strategies in User Feedback}

The reliability of annotations in user feedback depends on the annotation strategy, which can be categorized into expert-based, crowdsourced, and automated methods. 

Expert-based annotation, performed by domain specialists or trained annotators, provides high-quality labels through structured guidelines and domain knowledge. Studies on app review analysis, using Cohen's and Fleiss Kappa metrics~\cite{McHugh2012}, report \textit{moderate} ($0.41-0.60$) to \textit{substantial} ($0.61-0.80$) inter-annotator agreement, typically with Kappa values ranging from $0.60$ to $0.70$ for tasks such as feature extraction~\cite{Dabrowski2023} and sentiment analysis~\cite{Dabrowski2023,Riccosan2023}. However, expert annotation is costly and time-intensive, limiting scalability. Additionally, cognitive challenges in app review interpretation can limit the usefulness of human-annotated datasets to support automatic extraction, especially for descriptors such as review helpfulness, leading to \textit{slight} ($0.00-0.20$) agreement~\cite{Simmons2016}.%where only \textit{slight} agreement (Kappa $\leq 0.20$) is reached~\cite{Simmons2016}.  

Crowdsourced annotation utilizes non-expert contributors to label data at scale. Prior work has used crowdsourcing to validate unsupervised features and other descriptors in mobile app reviews~\cite{Motger2024,Alturaief2021}. Beyond app reviews, crowdsourcing is widely used in the context of RE~\cite{Hosseini2017,Vliet2020,Stanik2020}. 
While efficient, it often leads to \textit{moderate} agreement due to annotator subjectivity and domain unfamiliarity~\cite{Malpani2015,ALVARO2015280,Kim2018}. 

Automated methods leverage machine learning models or LLMs %(see Section~\ref{sec:background}) 
to generate annotations without human intervention~\cite{Hou2024}. While scalable, their reliability depends on model performance and training data quality. Some studies show comparable agreement levels between model-generated annotations and expert labels, particularly when fine-tuning encoder-based models on domain-specific data~\cite{Guo2023}. However, automated approaches lack human intuition, struggle with ambiguity, and may propagate biases from training data~\cite{Ashwin2023}.  
Consequently, expert annotation remains the gold standard.

In this study, we focus on expert-based labelling, guided by structured annotation guidelines, to ensure high-quality annotations. Due to the cognitive complexity of adapting emotions to app reviews, we exclude crowdsourced annotation. Finally, as LLMs demonstrate increasing effectiveness in text classification and annotation tasks~\cite{Hou2024}, we explore their potential as automated annotators in this context.
\section{Method}
\label{sec:method}

\subsection{Design}

% to better understand challenges, limitations [...] in emotion analysis of app reviews
% try to convey RQ1-RQ3 with RQ4
The goal of this research is \textbf{to identify and address the challenges and limitations of fine-grained emotion analysis in mobile app reviews}. Our study focuses on the adaptation of an emotion taxonomy, the analysis of the complexities of human annotation, and the reliability of LLMs for automated annotation. The following use cases illustrate how this fine-grained emotion analysis can support RE-related feedback gathering tasks by uncovering insights beyond traditional polarity-based analysis.

\begin{enumerate}[label=\textbf{UC\textsubscript{\arabic*.}}]
    \item \textbf{Categorizing feature-related feedback.} Identify distinct emotional responses tied to specific features, such as whether users struggle with usability, expect improvements, express confusion, or show appreciation.

    \item \textbf{Emotion-based user segmentation for release planning.} Segment user feedback by emotion types to understand how different user groups emotionally respond to recent updates or unmet expectations. This supports planning future releases that align better with user sentiment dynamics across versions or personas.
    
    \item \textbf{Prioritizing critical issues based on emotional risk.} Analyse high arousal reviews (e.g., anger over privacy violations, fear about data loss, or betrayal due to broken trust) to prioritize and address issues with higher potential for reputational or legal consequences.
\end{enumerate}

To achieve this goal and support the outlined use cases, we issue the following research questions (RQ): 

\begin{enumerate}[label=\textbf{RQ\textsubscript{\arabic*.}}]
    \item Which taxonomy of emotions is most suitable for annotating mobile app reviews?
    \item How can the selected taxonomy be effectively adapted to the specific context of app reviews?
    \item What challenges arise when humans manually annotate app reviews with emotion labels?
    \item How does LLM-based annotation compare to human annotation in emotion classification for app reviews?
\end{enumerate}

\begin{figure*}[t!]
  \centering
  \includegraphics[width=1\textwidth]{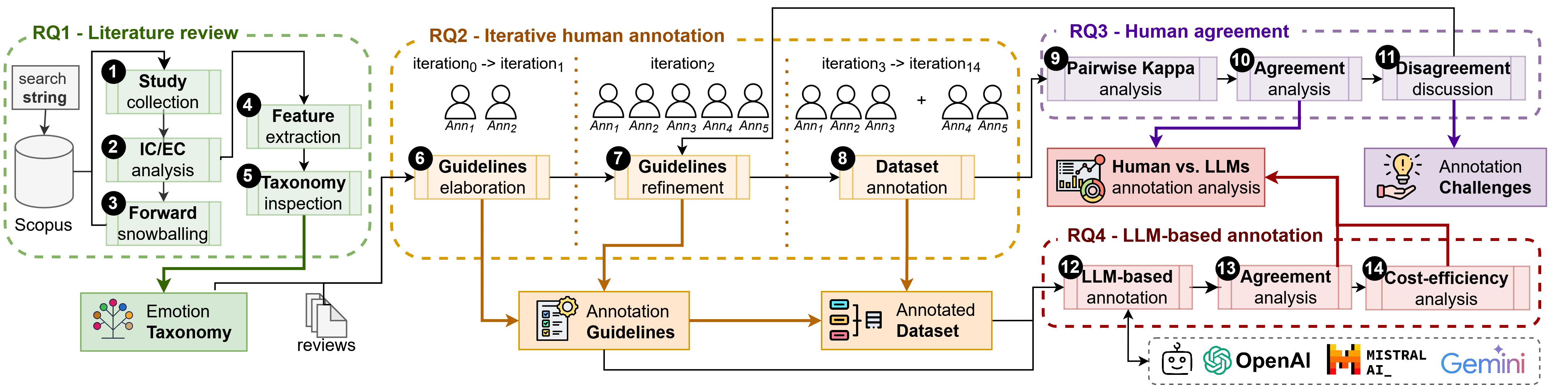}
  \caption{Research method.}
  \label{fig:method}
\end{figure*}

Figure~\ref{fig:method} illustrates our research design, including the steps involved in the resolution of each RQ. The figure shows how %, in the process of responding the research questions, 
our method has produced two actionable assets, ready to be used by the RE community: (1) \textit{annotation guidelines} to drive the manual process of annotating a set of app reviews using an emotion taxonomy; (b) an \textit{annotated dataset} containing 1,112 app reviews annotated with the emotion taxonomy following the annotation guidelines. 

To address \textbf{RQ\textsubscript{1}}, we conducted a \textbf{literature review} on emotion and sentiment analysis in user reviews. We extended the scope beyond mobile app reviews to software-related reviews for broader generalizability. This led to the selection of a suitable \textit{emotion taxonomy} as the foundation for annotation.

To address \textbf{RQ\textsubscript{2}}, we implemented an \textbf{iterative human annotation} process based upon \textit{annotation guidelines} using a subset of a publicly available app reviews dataset~\cite{Motger2024}. Feedback from each iteration served to refine the annotation guidelines, including definitions, instructions, and examples for identifying emotions in app reviews. These guidelines were used to generate the final \textit{annotated dataset}.

To address \textbf{RQ\textsubscript{3}}, we analysed \textbf{human agreement} during the annotation process. For each iteration, we computed the pairwise Kappa agreement among annotators and examined confusion matrices to detect label-specific interpretation conflicts. We analysed also the discussions held to resolve these conflicts, which served for clarifying biases or refining guidelines. This process identified key \textit{annotation challenges}, guiding the design of automatic emotion extraction tasks. % informing future automated emotion extraction approaches. % which could be addressed/leveraged/... in automatic emotion extraction X [tasks/methods/...]

To address \textbf{RQ\textsubscript{4}}, we designed an \textbf{LLM-based annotation} process, selecting and comparing state-of-the-art LLMs with advanced analytical capabilities. Our analysis focused on two dimensions: (1) cost-efficiency and (2) agreement, assessing inter-rater reliability between human and LLM annotations, as well as LLM prediction correctness against human ground truth. This resulted in a \textit{human vs. LLM annotation analysis}, laying the groundwork for LLM-based emotion extraction from app reviews and AI-based annotation in related tasks.
%This \textit{human vs. AI annotation analysis} provides validity to the annotation challenges reported in RQ\textsubscript{3}, as well as insights for the automatic extraction of emotions and the incorporation of AI assistants in annotation tasks.

Details of the methodology used in the four RQs follow.

\subsection{Literature Review}

We constructed our search string by integrating two key areas: \textit{emotion analysis} and \textit{software reviews}. For \textit{emotion analysis}, following Lin et al.'s approach~\cite{Lin2022}, we adopted the generic term \textit{emotion} to maximize potential matches. For \textit{software reviews}, we included multiple synonyms such as \textit{app}, \textit{application}, and \textit{API}, along with alternative phrases like \textit{user reviews}. This resulted in the following search string:\\

\vskip -9pt
\noindent{\texttt{\small("emotion*") AND ("app* review*" OR "software* review*" OR "user* review*" OR "application* review*" OR "api* review*")}}\\
\vskip -9pt

We selected Scopus for its broad coverage of high-quality, peer-reviewed research across disciplines~\cite{Martin-Martin2021}%. While Google Scholar offers the most extensive citation coverage, Scopus balances selectivity and breadth, indexing sources beyond computing, including social sciences and psychology. Compared to publisher-specific databases such as IEEE Xplore and ACM Digital Library, Scopus provides broader multidisciplinary coverage while maintaining reliable metadata quality and structured search functionalities.
%This makes it a suitable choice for capturing research at the intersection of emotion analysis and software reviews.
, structuring the literature review into the following steps:

\begin{itemize}

    \item \textbf{Step \step{1} -- Study collection.} We executed the defined search string in Scopus, exporting full references, metadata, and abstracts into a spreadsheet for further analysis.

    \item \textbf{Step \step{2} -- Inclusion and exclusion criteria (IC/EC).} %To ensure relevance, we applied predefined inclusion and exclusion criteria.
    Papers were included if they \textit{(i)} proposed, used, or analysed a multi-class emotion taxonomy, and \textit{(ii)} focused on software-related user reviews. Papers were excluded if they: \textit{(i)} lacked full-text access, \textit{(ii)} were not in English or Spanish, \textit{(iii)} were unrelated to opinion mining, or \textit{(iv)} focused solely on sentiment polarity. 

    \item \textbf{Step \step{3} -- Forward snowballing.} %For each study collection iteration, we assessed the need for additional snowballing.
    Backward snowballing was excluded as it prioritizes older studies, which may not align with the latest advancements in emotion analysis. Instead, we applied forward snowballing through Google Scholar to identify recent developments until saturation was reached, with no new relevant studies emerging.
    %Instead, we used forward snowballing to capture recent developments.

    \item \textbf{Step \step{4} -- Feature extraction.} For each selected study, we extracted key features, including taxonomy details (e.g., name, size, list of emotions), datasets (e.g., size, type, availability), emotion extraction methods (e.g., manual, machine learning, deep learning), and evaluation metrics (e.g., accuracy, precision, recall).

    \item \textbf{Step \step{5} -- Taxonomy inspection.} We analysed the identified emotion taxonomies both quantitatively and qualitatively. This assessment helped identify their potential benefits and limitations, ultimately guiding the selection of the most suitable taxonomy for this study. %\footnote{Our research method is taxonomy-agnostic, allowing replication studies to adopt a taxonomy that best fits their needs. The replication package is designed for generalization across different emotion taxonomies and annotation guidelines.}
    
\end{itemize}

\subsection{Iterative Human Annotation}
\label{sec:iterative-human-annotation}

Human annotation takes two key artifacts as input: the \textbf{emotion taxonomy} (derived from RQ\textsubscript{1}) and the \textbf{dataset of reviews} for annotation. We leveraged a dataset of user reviews from %Motger et al.~\cite{Motger2024}, which includes 
a multi-domain catalogue of popular mobile apps~\cite{Motger2024}, spanning 10 Google Play categories. Additionally, it provides annotations for 198 distinct app features (e.g., \textit{instant messaging}, \textit{video sharing}, \textit{note-taking}, \textit{GPS navigation}), enabling future research in feature-based emotion analysis. We randomly selected up to 10 sentences from reviews mentioning each feature\footnote{Some features in the dataset had fewer than 10 reviews mentioning them.}, resulting in a total of 1,412 reviews. These were shuffled and divided into 15 subsets (100 reviews each, 12 for the latter), labelled from $iteration_0$ to $iteration_{14}$. Annotations were performed by five human annotators ($Ann_1$ to $Ann_5$), all co-authors of this paper, as follows:

\begin{itemize}
    \item \textbf{Step \step{6} -- Guidelines elaboration.} Based on the literature review (RQ\textsubscript{1}), $Ann_1$ drafted the initial version of the annotation guidelines. These guidelines include: \textit{(i)} formal definitions of each emotion adapted to the context of mobile app reviews, %\textit{(ii)} domain-specific adaptations for mobile app reviews, 
    \textit{(ii)} real examples of app review sentences for each emotion, and \textit{(iii)} detailed instructions on the annotation process. Annotation was defined at the sentence level, with emotions assigned atomically to individual sentences. Annotators could refer to full reviews for context and ambiguity resolution. The guidelines were refined during the first two iterations. In $iteration_0$, $Ann_1$ and $Ann_2$ independently labelled a shared subset of sentences without prior discussion, allowing ambiguities and gaps in the guidelines to surface. %These were addressed in subsequent iterations until substantial agreement was achieved (see Section~\ref{sec:agreement-analysis}).

    \item \textbf{Step \step{7} -- Guidelines refinement.} Following the initial guidelines draft, all five annotators participated in a joint iteration ($iteration_2$). Involving multiple annotators ensured diverse perspectives, helping to uncover defects, inconsistencies, ambiguities, and potential threats to validity in the annotation process. After annotation, we conducted a dedicated meeting to systematically review disagreements, linking each to a concrete action point for improving the annotation guidelines (e.g., adding examples, refining ambiguous vocabulary, clarifying criteria for distinguishing emotions).
        
    \item \textbf{Step \step{8} -- Dataset annotation.} Once the guidelines were stabilized, we iterated over the remaining 12 subsets of reviews ($iteration_3 \rightarrow iteration_{14}$) to create the final dataset. Three annotators ($Ann_1,Ann_2,Ann_3$) were designated as \textit{main annotators} and two ($Ann_4,Ann_5$) as \textit{secondary annotators}. Each iteration included two main annotators and one secondary annotator, rotating across all possible pairings to systematically identify and address inconsistencies. The 12 iterations were grouped into four batches (2, 4, 4, and 2 iterations, respectively). Each annotator recorded the time spent reading the guidelines and completing each iteration. %After each batch, we conducted agreement analysis meetings to refine the guidelines, resolve disagreements, and improve the annotation process for subsequent batches (see Section~\ref{sec:agreement-analysis}). 
    Each emotion was assessed independently, meaning that a given sentence can be annotated with multiple emotions. This possibility was used with caution, i.e. only when the sentence contained sub-sentences expressing different emotions. The final annotation retained emotion labels that at least two of the three annotators involved in the iteration agreed upon.

\end{itemize}

\subsection{Human Agreement}
\label{sec:agreement-analysis}

After each iteration, we measured annotator agreement, analysed disagreements, and refined the guidelines to improve annotation consistency. This process was structured as follows:

\begin{itemize}
    \item \textbf{Step \step{9} -- Pairwise Kappa analysis.} We used Cohen's Kappa to assess pairwise agreement between annotators in each iteration. Unlike Fleiss' Kappa, pairwise Cohen’s Kappa allows us to monitor individual interpretation conflicts and cognitive biases, helping to identify challenges in emotion classification from user reviews. %Additionally, several studies claim that Fleiss' Kappa an Cohen's Kappa yield similar results~\cite{Mitani2017,FLETCHER2011341}. 
    To further analyse disagreement patterns, we also generated confusion matrices for each pair of annotators.
    
    \item \textbf{Step \step{10} -- Agreement analysis.} As a quality control measure, we set a minimum agreement threshold of Cohen's Kappa $\geq 0.60$ (i.e., \textit{substantial} agreement) for each pair of annotators per iteration. Additionally, we qualitatively analysed the confusion matrices from the previous step to identify recurring disagreement patterns (e.g., frequently confused emotions, systematic biases, or inconsistencies in annotator tendencies).
    
    \item \textbf{Step \step{11} -- Disagreement discussion.} Using Cohen's Kappa and confusion matrices as input, we held dedicated meetings at the end of each batch to analyse major disagreement patterns. Each pattern was linked to a specific action point, akin to those proposed in Step \step{7}. At this stage, we aimed to minimize guideline modifications to maintain consistency with previous annotations. Changes were kept limited and comparable in scope to ensure that the guidelines remained generalizable across different datasets and app domains while avoiding overstatements.

\end{itemize}

\subsection{LLM-based Annotation}
\label{sec:llm-based-agreement}

%After human annotation, we %used the annotated dataset to 
%evaluated the performance of LLM-based agents instructed with the generated annotation guidelines. This process was designed as follows:
After annotation, we evaluated LLM agents using the generated guidelines as follows:

\begin{itemize}
    \item \textbf{Step \step{12} -- LLM-based annotation.} We evaluated the performance of three advanced LLMs with API access: GPT-4o\footnote{\href{https://platform.openai.com/docs/models/gpt-4o}{https://platform.openai.com/docs/models/gpt-4o}}, Mistral Large 2\footnote{\href{https://mistral.ai/en/news/mistral-large-2407}{https://mistral.ai/en/news/mistral-large-2407}}, and Gemini 2.0 Flash\footnote{\href{https://deepmind.google/technologies/gemini/flash/}{https://deepmind.google/technologies/gemini/flash/}}. For each, we created an LLM assistant with a system prompt embedding the annotation guidelines with additional input-output formatting instructions. We tested the models under three temperature settings: high (1), mid (0.5), and low (0). This experimental setup enables a more comprehensive evaluation of LLM performance while aligning LLM challenges with those encountered in human annotation (RQ\textsubscript{3}). To balance efficiency and avoid performance degradation from excessively long prompts, the dataset was processed in batches of 10 reviews. Each assistant was run three times on the full dataset.

    \item \textbf{Step \step{13} -- Agreement analysis.} %Using the average results from three LLM annotation runs, %we quantitatively compared the cost-efficiency of human and LLM-based annotation. Additionally,
    We assessed the average pairwise agreement between humans and LLM assistants using Cohen's Kappa. Additionally, treating the human agreement as ground truth, we evaluated LLM performance in terms of precision, recall, and F-measure. 
        
    \item \textbf{Step \step{14} -- Cost-efficiency analysis.} Using average results from three annotation runs for each LLM, we measured total token usage, including input and completion output tokens, along with the associated API cost (€). We also recorded execution times for the automated annotation process. These results were then compared to a human cost-efficiency analysis, factoring in personnel costs (€) and the time required for annotation iterations.

\end{itemize}
\section{Results}
\label{sec:results}

\subsection{Emotion Taxonomy (RQ\textsubscript{1})}
\label{sec:res-literature-review}

\begin{figure}[t] \centering \includegraphics[width=\linewidth]{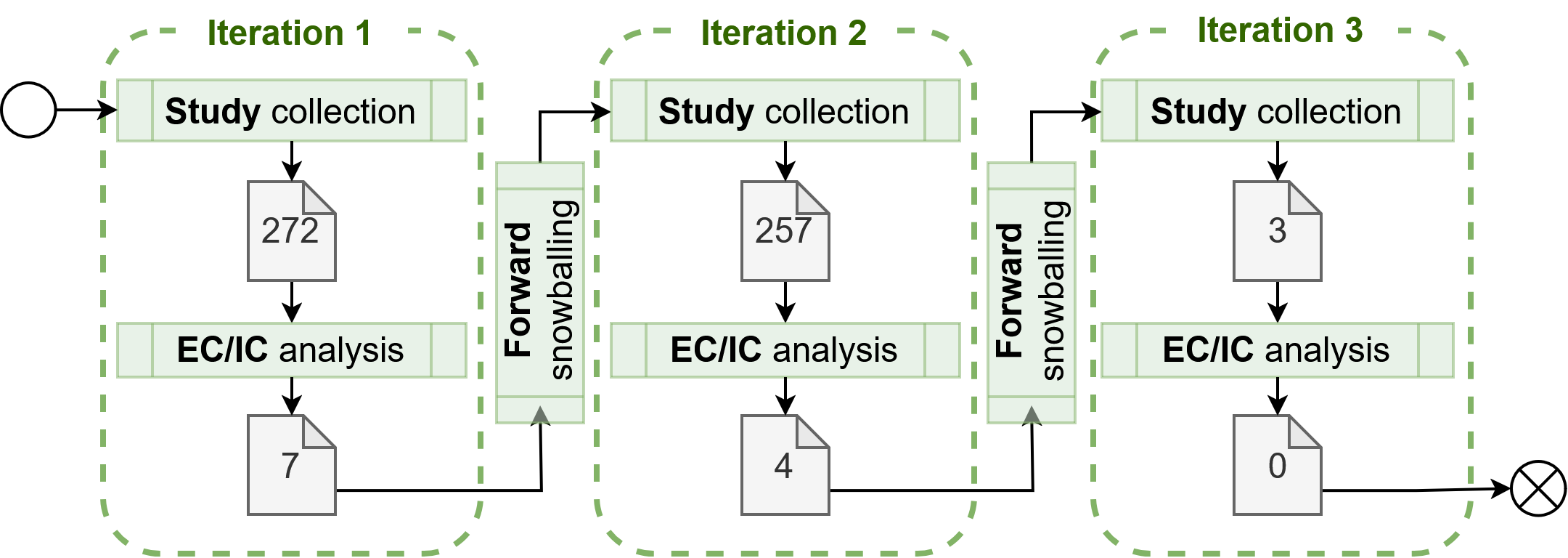} \caption{Results from the literature review.} \label{fig:literature-review} \end{figure}

Figure~\ref{fig:literature-review} reports the results of the literature review\footnote{Study collection on iteration 1 was conducted on June 1st, 2024.}, which identified 11 studies employing multi-class emotion taxonomies in the context of software reviews.% (7 in the first iteration, 4 in the second).

No single taxonomy has been universally adopted for emotion analysis in software reviews. Three studies~\cite{Pawełoszek2023,Shaheen2019,Singh2022101929} utilize or adapt Plutchik's wheel of emotions~\cite{Plutchik2001}, which defines eight basic emotions -- \textit{Joy}, \textit{Trust}, \textit{Fear}, \textit{Surprise}, \textit{Sadness}, \textit{Disgust}, \textit{Anger}, and \textit{Anticipation} -- while also modeling relationships between emotions and their arousal levels (i.e., intensity of emotional experience).
Two studies~\cite{Keertipati2016,Riccosan2023} employ variations of Parrott's taxonomy~\cite{Parrott2001}, which identifies six basic emotions. Five of these emotions (\textit{Joy}, \textit{Fear}, \textit{Surprise}, \textit{Sadness}, and \textit{Anger}) align with Plutchik's, while \textit{Love} is introduced as a distinct category. 
One study~\cite{Malgaonkar2019} applies Ekman's taxonomy~\cite{Ekman1992}, which includes \textit{Happy}, \textit{Sadness}, \textit{Surprise}, \textit{Fear}, and \textit{Anger}, overlapping with both Plutchik’s and Parrott’s models. Additionally, it incorporates \textit{Disgust}, consistent with Plutchik’s classification.
Another study~\cite{Cabellos2022} utilizes Liew and Turtle's taxonomy, which defines 28 fine-grained emotions, including \textit{Admiration}, \textit{Doubt}, \textit{Pride}, and \textit{Jealousy}, providing a more granular categorization of emotional expressions in software reviews. 
Finally, four studies~\cite{Valenzuela2021,Gao2018,Myers2019,Savarimuthu2023} employed custom emotion taxonomies. These approaches included reusing emotions from existing taxonomies~\cite{Myers2019}, adapting taxonomies to align with the syntax required by a third-party emotion extraction tool~\cite{Gao2018,Savarimuthu2023}, or defining a domain-specific set of emotions~\cite{Valenzuela2021}.

\begin{figure}[h]
  \centering
  \includegraphics[width=1\linewidth]{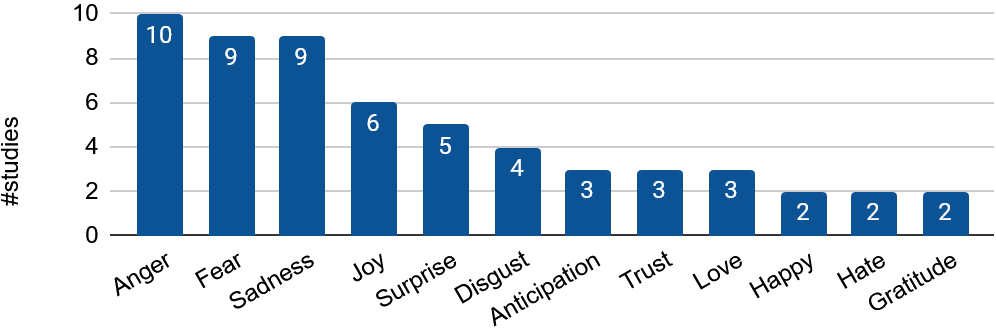}
  \caption{Distribution of emotions in the literature review (including only those that appear in more than one study).}
  \label{fig:emotions}
\end{figure}

We collected all identified emotions and applied minimal normalization, which involved standardizing syntax by converting words to their root forms. Using normalized data, we generated Figure~\ref{fig:emotions} to illustrate the frequency of emotions studied in software reviews. Based on this analysis, we chose Plutchik's Wheel of Emotions not only because it is the most used in the papers found, but also for the following reasons. First, its eight basic emotions are among the nine most studied, the only exception being \textit{Love}. Second, Plutchik’s model defines opposite emotion pairs and incorporates contiguous emotions, making it easier to compare emotions while allowing the identification and discussion of reviews that may fall between two emotions. Finally, emotions such as \textit{Anticipation} -- unique to Plutchik’s model -- can provide insights into user expectations, particularly in exploring new features, a common focus in mobile app feedback analysis~\cite{Maalej2015,Guzman2014}.

\subsection{Annotation Guidelines and Annotated Dataset (RQ\textsubscript{2})}
\label{sec:guidelines-dataset}

The outcome of RQ\textsubscript{2} consists of two artifacts: (1) the \textit{annotation guidelines}, and (2) the \textit{annotated dataset}.

\begin{table*}[t]
\centering
\setlength{\tabcolsep}{3.5pt}
\caption{Summary of emotion annotation guidelines. Includes number of annotated sentences per label (\#).}
\begin{tabular}{lcp{0.18\linewidth}p{0.32\linewidth}p{0.30\linewidth}} 
\toprule
\textbf{Label} & \textbf{\#} & \textbf{Expresses...} & \textbf{Contains...} & \textbf{Example} \\ \midrule
\joy & 372 & excitement; pleasure; possibility; positiveness & app praise; features or characteristics appreciated by users & \textit{``Excellent task and list management tool''} \\ %\hline
\sadness & 303 & disappointment; loss; heaviness; severity & negative user experiences; reports of bugs or faults & \textit{``Just that RCS/Chat function has regular failures unfortunately.''} \\ %\hline
\anticipation & 182 & curiosity; consideration; alertness; exploration & inquiries about app changes; feature requests; questions about uncertain aspects & \textit{``I would love a barcode scanner to be included.''} \\ %\hline
\trust & 151 & acceptance; connection; safeness; warmth & positive user experiences; satisfaction with goals or needs; privacy and security observations & \textit{``One of the best note-taking I've tried so far''} \\ %\hline
\surprise & 54 & shocked; unexpected; unpredictability & unexpected user issues; unknown events or situations & \textit{``Why you made char limit in bookmark name and description???''} \\ %\hline
\disgust & 49 & lack of trust; rejection; bitterness; refusal & explicit rejections of the app; claims that the app is unsuitable for its purpose & \textit{``I find this app frustrating, will be using a running watch from now on''} \\ %\hline
\anger & 27 & fury; rage; fierceness; hate & hate speech; insults; extremely negative vocabulary & \textit{``I hate that it disconnects if the app is not running in the background''} \\ %\hline
\fear & 24 & stress; anxiety; agitation; scared & complaints about trust issues; requests for help with frustration or anxiety & \textit{``People are adding fake members via python by scrapping to make their group look huge.''} \\ %\hline
\neutral & 88 & nothing & objective content; no emotions regarding current app state & \textit{``Call recording is disabled.''} \\ %\hline
\reject & 22 & unknown & non-English text; noisy content & \textit{``Video chat ki bii suvida honi chahiy''} \\ \bottomrule
\end{tabular}
\label{tab:emotion-guidelines}
\end{table*}

\subsubsection{Annotation guidelines} 
A 10-page document including formal definitions of each of the Plutchnik's eight primary emotions adapted to the mobile app review domain, alongside 48 annotated review sentence examples. Table~\ref{tab:emotion-guidelines} summarizes (partially) the content of these guidelines, which are fully detailed in our replication package. In addition to Plutchnik's emotions, during this process we elicited two additional labels for the annotation task: \neutral, restricted to sentences reflecting purely objective content without expressing any particular emotion linked to the current state of the app; and \textit{\reject}, restricted to sentences that cannot be interpreted from a linguistic standpoint. % (e.g., non-English reviews, noisy content, indecipherable meaning...).

\subsubsection{Annotated dataset}
A dataset of 1,272 annotated labels, including emotions, neutral, and rejected cases, assigned to 1,112 sentences from distinct reviews. Table~\ref{tab:emotion-guidelines} includes the distribution for each emotion in the final dataset, in addition to 88 neutral sentences and 22 rejected sentences.

\subsection{Annotation Challenges (RQ\textsubscript{3})}
\label{sec:res-challenges}

\begin{figure}[h]
  \centering
  \includegraphics[width=\linewidth]{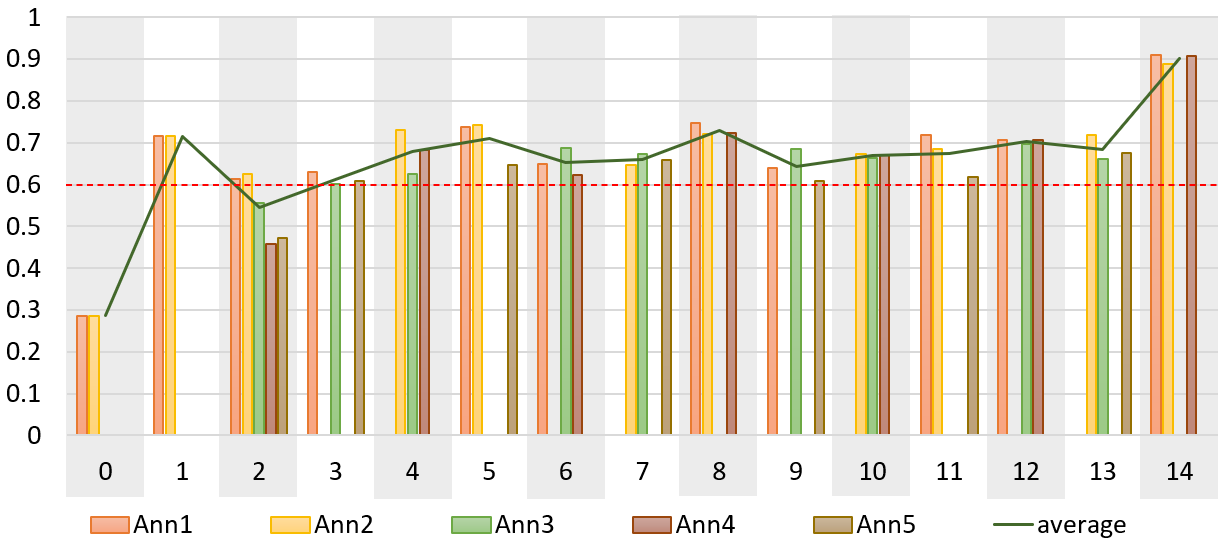}
  \caption{Evolution of the average Cohen's Kappa agreement across iterations.}
  \label{fig:agreement}
\end{figure}

Figure~\ref{fig:agreement} summarizes the evolution of the average Cohen's Kappa agreement for each annotator across iterations following the process described in Sections~\ref{sec:iterative-human-annotation} and~\ref{sec:agreement-analysis}. 
%ho resumeixo una mica

After the initial version of the guidelines produced by $Ann_1$ and $Ann_2$ in $iteration_1$, the team inter-rater agreement grew progressively from \textit{moderate} at the start of the process (average Cohen’s Kappa of $0.52$ in $iteration_2$) to \textit{substantial} at the end (average Cohen's Kappa of $0.69$ for all iterations involving the dataset annotation). Notably, all pairwise agreements from $iteration_3$ onward remained above the \textit{substantial} agreement threshold (dotted red line). Fluctuations in early iterations suggest ongoing adjustments into the guidelines, while later stability reflects the effectiveness of iterative refinement. The final iteration showed higher agreement but involved only 12 reviews, making it less representative. Complementarily, we also analysed average Cohen's Kappa scores for individual emotion labels. Results show that \joyA, \anticipationA, and \reject labels achieved the highest agreement, while \surpriseA, \angerA, and \disgust exhibited the lowest consistency. Expanded label-specific agreement is available in the replication package. %Finally, $Ann_1$ and $Ann_2$ exhibited higher agreement, likely due to their involvement in the initial guideline elaboration.

%In \textit{iteration\_0}, when only the initial version of the annotation guidelines was available, the agreement was fair ($0.27$). In \textit{iteration\_1}, following discussion (Steps \step{4} $\rightarrow$ \step{6}), agreement increased substantially up to $0.72$. Once a stable version of the guidelines was established, all annotators participated in \textit{iteration\_2}. At this stage, using only the guidelines, the average pairwise Cohen's Kappa agreement was moderate ($0.52$). After further discussions and refinement of the guidelines, we conducted four dataset annotation batches over the next 12 iterations. All subsequent iterations achieved substantial agreement, with an average pairwise Cohen’s Kappa of $0.69$.

During this process, we experienced a number of challenges in adapting Plutchik's taxonomy to mobile app reviews and developed mitigation strategies, either incorporated into the guidelines or applied during annotation. %Additionally, we propose design guidelines for researchers using annotated datasets for automatic emotion extraction. 
We summarize the key outcomes below:

\begin{itemize}
    \item\textbf{Challenge 1. Defining boundaries between contiguous emotions.}  
    Contiguous emotions in Plutchik's taxonomy share overlapping attributes, leading to consistent disagreements among annotators. For instance, in the sentence \textit{``I just love this notebook''}, %(\textit{iteration\_7}),
    $Ann_2$ and $Ann_3$ labelled it as \joyA, reflecting app appraisal, while $Ann_5$ assigned \trustA, interpreting it as a personal involvement. Similarly, in \textit{``[...] the app is saying I need to keep signing in and my notebooks aren't retrievable''}, % (\textit{iteration\_2}), 
    all annotators marked it as \sadnessA, emphasizing disappointment, while $Ann_4$ labelled it as \disgustA, focusing on rejection.
    %For instance, \joy and \trust both convey positive sentiment (e.g., \textit{``I just love this notebook''}), but \joy reflects app appraisal, while \trust implies personal involvement. Similarly, \sadness and \disgust are both negative (e.g., \textit{``[...] the app is saying I need to keep signing in and my notebooks aren't retrievable.''}) but differ in disappointment versus rejection.  
    To address this, we refined the guidelines with explicit disambiguation criteria for conflicting emotion pairs reporting the lowest label-specific agreement. 

    \item\noindent\textbf{Challenge 2. Addressing mixed or conflicting emotions in a single sentence.}  
    A sentence may express multiple, potentially conflicting emotions, such as \joy and \sadness (e.g., \textit{``I really appreciate this app but lately I've been having issues with the cloud sync.''}) or \sadness and \anger (e.g., \textit{``It wasn't what I needed, and I absolutely HATE that they don't even tell you the timers are for premium memberships only.''}). This complexity challenges the assumption that emotions are mutually exclusive.  
    To address this, we allowed multiple emotions to be assigned to a single sentence, ensuring that mixed - or even conflicting - emotions are properly captured without enforcing artificial exclusivity.
    %To address this, we revised our guidelines to treat all emotions equally, without hierarchical distinctions. 
    
    \item\textbf{Challenge 3. Establishing thresholds for subtle emotional intensity (arousal).}  
    Implicit emotional expressions may be overlooked or exaggerated due to subjective interpretation. For instance, \sadness or disappointment can be difficult to assess (e.g., \textit{``No calendar sync''}), and \surprise may depend on whether an event was truly unexpected (e.g., \textit{``The app decided to add a folder to my gallery [...]''}). 
    To address this, we refined our guidelines with additional examples to clarify when an emotion's intensity meets the labeling threshold, reducing both underreporting and overinterpretation. 
    
   \item\textbf{Challenge 4. Handling lack of context and linguistic ambiguities in sentence-level emotion extraction.}  
    Eliciting emotions at the sentence level is challenging due to missing contextual information. This could potentially lead to inconsistencies, as identical sentences yield different emotions depending on their context, particularly when linguistic ambiguities (e.g., pronouns, acronyms, or elliptical subjects) were present. To address this, annotators were instructed to label individual sentences, consulting the full review only to resolve linguistic ambiguities. 
    
\end{itemize}

\subsection{Human vs. LLM Annotation Analysis (RQ\textsubscript{4})}

As described in Section~\ref{sec:method}, annotation analysis between human and LLM-based performance focuses on two main dimensions: agreement and cost-efficiency analysis.

\begin{figure}[h]
  \centering
  \includegraphics[width=\linewidth]{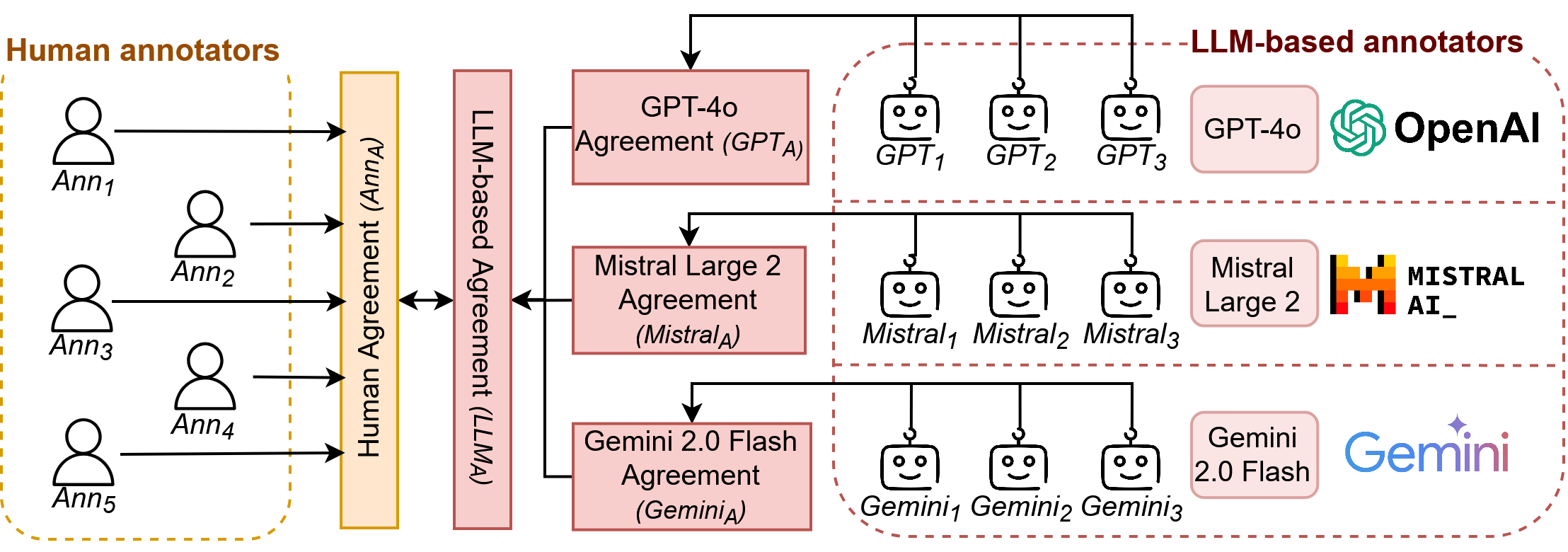}
  \caption{Evaluation of human agreement vs. LLM-based agreement}
  \label{fig:agreement-design}
\end{figure}

\subsubsection{Agreement Analysis}

Figure~\ref{fig:agreement-design} illustrates the evaluation setup, comparing human agreement ($Ann_A$) with LLM-based agreement ($LLM_A$). For each LLM included in this study, we conducted three annotation runs, deriving an agreement annotation using the same criteria applied to human agreement (see Section~\ref{sec:iterative-human-annotation}). We then computed the agreement among individual LLMs (GPT-4o, Mistral Large 2, Gemini 2.0 Flash) to obtain the overall LLM agreement ($LLM_A$), which we compared against human agreement ($Ann_A$). Since the lowest temperature setting (0) yielded the best results, we limit the reported results to this setting.

\begin{figure}[h]
  \centering
  \includegraphics[width=\linewidth]{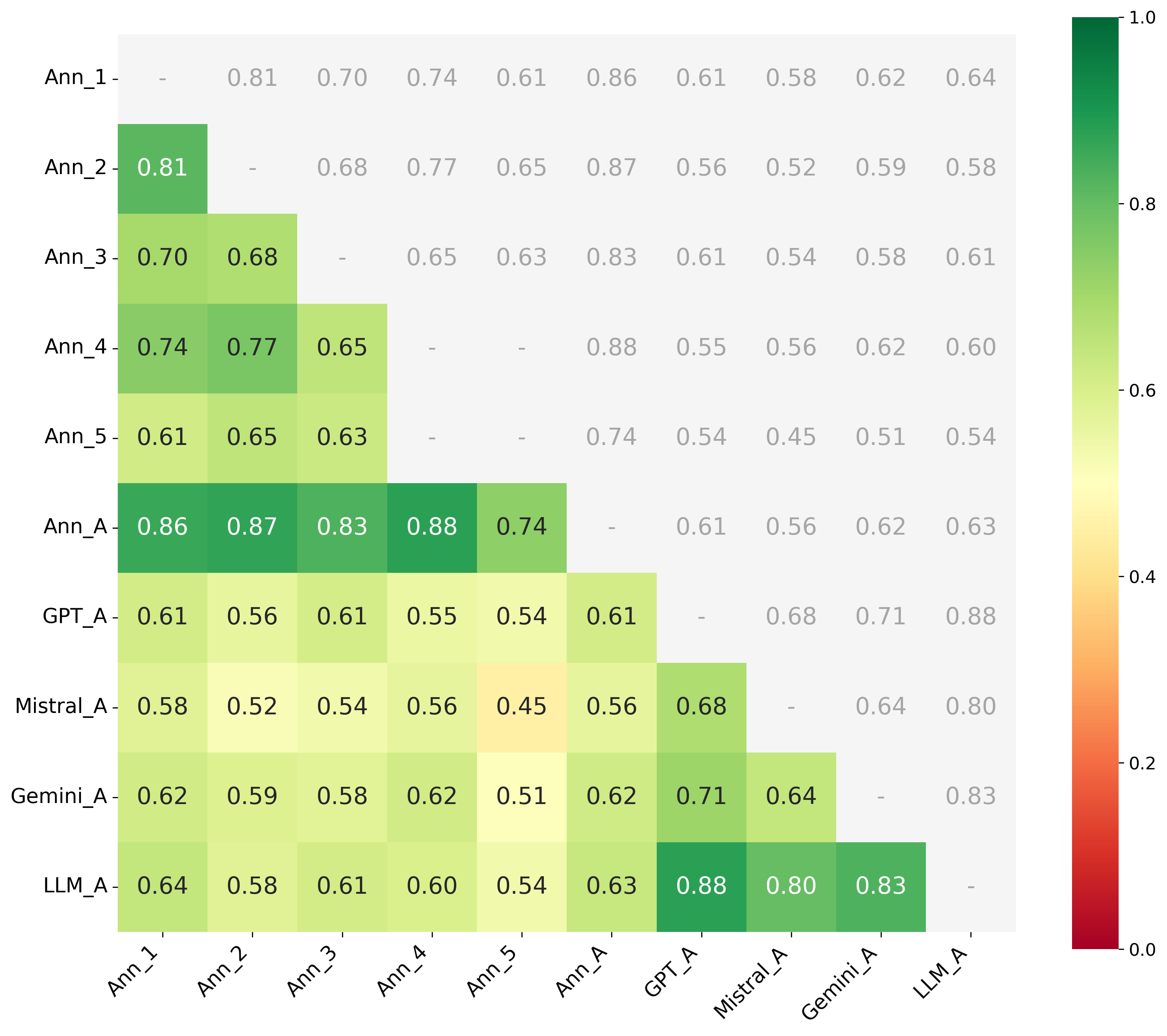}
  \caption{Inter-rater pairwise agreement (human and LLM-based)}
  \label{fig:agreement-heatmap}
\end{figure}

%LESS VERBOSE
Figure~\ref{fig:agreement-heatmap} reports the inter-rater pairwise Cohen's Kappa agreement between all annotators, including five humans\footnote{Secondary annotators $Ann_4$ and $Ann_5$ never participated in the same iteration of the final dataset annotation ($iteration_3$ to $iteration_{14}$).} 
($Ann_i$) and LLMs ($GPT_A$, $Mistral_A$, $Gemini_A$). It also includes agreement between each annotator and the overall human ($Ann_A$) and LLM-based ($LLM_A$) agreements. Agreement between different runs from the same LLM is excluded from this analysis, as they exhibited near-perfect consistency (pairwise Cohen's Kappa agreement was always $\geq0.87$). 
%On average, humans show similar agreement among themselves ($0.69$) as LLMs . 
On average, pairs of humans show an agreement ($0.69$) that is similar to the agreement shown by pairs of LLMs ($0.68$). However, LLM annotations ($GPT_A$, $Mistral_A$, $Gemini_A$) exhibit moderate agreement with human annotations ($0.56-0.62$), indicating some deviation from human judgment. 
Among individual models, Gemini 2.0 Flash ($Gemini_A$) aligns more closely with human agreement ($Ann_A$), although LLM agreement ($LLM_A$) is slightly higher. This suggests that combining multiple LLM instances helps mitigate biases, similar to how human annotators vary in judgment. GPT-4o ($GPT_A$) shows the highest consistency with overall LLM agreement ($LLM_A$), while Mistral and Gemini introduce more variability across different runs. %, making their annotations less stable across different runs.

\begin{table}[t]
\caption{LLM correctness vs. human agreement ($Ann_A$)}
\label{tab:correctness}
\centering{%
%\begin{tabularx}{\linewidth}{lXXX}
\begin{tabularx}{\linewidth}{l>{\centering\arraybackslash}X>{\centering\arraybackslash}X>{\centering\arraybackslash}X}
\toprule
\textbf{Annotator}             & \textbf{Precision} & \textbf{Recall}  & \textbf{F1}      \\ \midrule
GPT\textsubscript{i}         & 68.77\%            & 64.50\%          & 66.57\%          \\
GPT\textsubscript{A}            & 69.41\%   & 64.92\%          & 67.09\%          \\ \midrule
Mistral\textsubscript{i}   & 62.26\%            & 61.13\%          & 61.69\%          \\
Mistral\textsubscript{A}   & 63.36\%            & 61.71\%          & 62.52\%          \\ \midrule
Gemini\textsubscript{i}  & 67.44\%            & 67.46\%          & 67.45\%          \\
Gemini\textsubscript{A}  & 67.48\%            & \cellcolor{Green!30}\textbf{67.59\%} & 67.54\% \\ \midrule
LLM\textsubscript{A}  & \cellcolor{Green!30}\textbf{70.68\%}            & 66.81\% & \cellcolor{Green!30}\textbf{68.69\%} \\ 
\bottomrule
\end{tabularx}%
}
\end{table}

Using human agreement ($Ann_A$) as ground-truth, Table~\ref{tab:correctness} reports the average precision, recall and F1 metrics %\footnote{We exclude accuracy due to the large number of true negatives.} 
of: individual LLM runs ($GPT_i$, $Mistral_i$, $Gemini_i$); agreement between LLM runs ($GPT_A$, $Mistral_A$, $Gemini_A$); and agreement between all LLMs ($LLM_A$). These results reinforce the idea that agreement between multiple LLM instances ($LLM_A$) is the best setting, reporting the highest F1 measure. However, the highest recall is reported by $Gemini_A$. This also opens the need to analyse alternative quality metrics, properly weighting precision and recall based on task criticality and the cost of missing relevant emotion labels~\cite{Berry2021}.

\begin{table}[h]
\caption{Human individual correctness vs. human agreement ($Ann_A$)}
\label{tab:correctness-human}
\centering{%
%\begin{tabularx}{\linewidth}{lXXX}
\begin{tabularx}{\linewidth}{l>{\centering\arraybackslash}X>{\centering\arraybackslash}X>{\centering\arraybackslash}X}
\toprule
\textbf{Annotator}             & \textbf{Precision} & \textbf{Recall}  & \textbf{F1}      \\ \midrule
Ann\textsubscript{1}         & 86.96\%            & 88.76\%          & 87.86\%          \\
Ann\textsubscript{2}         & \cellcolor{Green!30}\textbf{88.66\%}            & \cellcolor{Green!30}\textbf{88.89\%}          & \cellcolor{Green!30}\textbf{88.78\%}          \\
Ann\textsubscript{3}         & 86.14\%            & 84.84\%          & 85.49\%          \\
Ann\textsubscript{4}         & 85.24\%            & 83.08\%          & 84.16\%          \\
Ann\textsubscript{5}         & 81.97\%            & 77.73\%          & 79.85\%          \\
\midrule
Ann\textsubscript{i}         & 85.79\%            & 84.66\%          & 85.23\%          \\
\bottomrule
\end{tabularx}%
}
\end{table}

Complementarily, and using the same criteria, Table~\ref{tab:correctness-human} shows that human annotators are highly consistent with one another, with F1 scores ranging narrowly from 79.85\% to 88.78\% and an overall average of 85.23\%. Despite having different roles in the annotation process (e.g., primary vs. secondary), all annotators achieve comparably high precision and recall, indicating strong alignment with the agreed-upon annotation guidelines. This consistency reinforces the reliability of the human-created ground truth and highlights the effectiveness of the annotation guidelines and calibration process in minimizing individual biases.

% \begin{table}[h]
% \caption{LLM correctness vs. human ($Ann_A$)}
% \label{tab:correctness}
% \centering{%
% \begin{tabularx}{\linewidth}{lXXXX}
% \toprule
% \textbf{Annotator}            & \textbf{Accuracy} & \textbf{Precision} & \textbf{Recall}  & \textbf{F1}      \\ \midrule
% GPT\textsubscript{i}         & 88.05\%           & 68.77\%            & 64.50\%          & 66.57\%          \\
% GPT\textsubscript{A}           & \cellcolor{Green!30}\textbf{92.64\%}  & 69.41\%   & 64.92\%          & 67.09\%          \\ \midrule
% Mistral\textsubscript{i}   & 85.91\%           & 62.26\%            & 61.13\%          & 61.69\%          \\
% Mistral\textsubscript{A}   & 86.31\%           & 63.36\%            & 61.71\%          & 62.52\%          \\ \midrule
% Gemini\textsubscript{i}  & 88.19\%           & 67.44\%            & 67.46\%          & 67.45\%          \\
% Gemini\textsubscript{A}  & 88.29\%           & 67.48\%            & \cellcolor{Green!30}\textbf{67.59\%} & 67.54\% \\ \midrule
% LLM\textsubscript{A}  & 88.80\%           & \cellcolor{Green!30}\textbf{70.68\%}            & 66.81\% & \cellcolor{Green!30}\textbf{68.69\%} \\ 
% \bottomrule
% \end{tabularx}%
% }
% \end{table}

Table~\ref{tab:correctness-emotion} presents correctness metrics for each annotation label defined in Table~\ref{tab:emotion-guidelines}, revealing substantial differences in annotation performance across labels. The best-performing classes —- \joyA, \sadnessA, \anticipationA, and \reject —- exhibit high recall and balanced precision, suggesting that these are easier to distinguish based on the guidelines used by the LLM annotator. Notably, these labels also show the highest human agreement (RQ\textsubscript{3}), reinforcing their clearer annotation boundaries. Conversely, \surpriseA, \disgustA, \angerA, and \fear show the lowest performance, both in terms of correctness and label-specific human agreement, with frequent misclassifications. Additionally, \anger has relatively high recall but very low precision, indicating it is often over-predicted, whereas \surprise suffers from poor recall, suggesting the model struggles to identify it. 
These discrepancies highlight the challenges LLMs face in distinguishing subtle or less frequently expressed emotions (\textit{Challenge 3} in RQ\textsubscript{3}). Overall, these findings emphasize the need for refined annotation guidelines and model adaptations, such as class-specific confidence thresholds or multi-label classification approaches. 

\begin{table}[t]
\caption{LLM agreement ($LLM_A$) correctness per label}
\label{tab:correctness-emotion}
\centering
\begin{tabularx}{\linewidth}{l>{\centering\arraybackslash}X>{\centering\arraybackslash}X>{\centering\arraybackslash}X>{\centering\arraybackslash}X}
\toprule
\textbf{Label} & \textbf{Precision} & \textbf{Recall} & \textbf{F1} \\ \midrule
\joy          & \colorize{73.70} & \colorize{68.55} & \colorize{71.03} \\
\sadness      & \colorize{82.48} & \colorize{74.59} & \colorize{78.34} \\
\anticipation & \colorize{78.57} & \colorize{90.66} & \colorize{84.18} \\
\trust        & \colorize{70.91} & \colorize{51.32} & \colorize{59.54} \\
\surprise     & \colorize{41.38} & \colorize{22.22} & \colorize{28.92} \\
\disgust      & \colorize{42.86} & \colorize{42.00} & \colorize{42.42} \\
\anger        & \colorize{35.42} & \colorize{62.96} & \colorize{45.33} \\
\fear         & \colorize{38.10} & \colorize{33.33} & \colorize{35.56} \\
\neutral      & \colorize{48.15} & \colorize{59.09} & \colorize{53.06} \\
\reject       & \colorize{94.44} & \colorize{77.27} & \colorize{85.00} \\ \bottomrule
\end{tabularx}
\end{table}

\begin{table}[h]
\caption{Average human individual ($Ann_i$) correctness per label}
\label{tab:correctness-emotion-human}
\centering
\begin{tabularx}{\linewidth}{l>{\centering\arraybackslash}X>{\centering\arraybackslash}X>{\centering\arraybackslash}X>{\centering\arraybackslash}X}
\toprule
\textbf{Label} & \textbf{Precision} & \textbf{Recall} & \textbf{F1} \\ \midrule
\joy          & \colorize{90.99} & \colorize{91.37} & \colorize{91.18} \\
\sadness      & \colorize{89.36} & \colorize{82.67} & \colorize{86.02} \\
\anticipation & \colorize{88.71} & \colorize{88.72} & \colorize{88.71} \\
\trust        & \colorize{85.23} & \colorize{81.37} & \colorize{83.30} \\
\surprise     & \colorize{64.07} & \colorize{74.56} & \colorize{69.31} \\
\disgust      & \colorize{68.98} & \colorize{72.62} & \colorize{70.80} \\
\anger        & \colorize{69.07} & \colorize{77.21} & \colorize{73.14} \\
\fear         & \colorize{74.15} & \colorize{78.50} & \colorize{76.32} \\
\neutral      & \colorize{78.41} & \colorize{76.06} & \colorize{77.24} \\
\reject       & \colorize{82.00} & \colorize{89.46} & \colorize{85.73} \\ \bottomrule
\end{tabularx}
\end{table}

Table~\ref{tab:correctness-emotion-human} shows the average correctness of individual human annotators per emotion label. While humans also find emotions like \surpriseA, \disgustA, \angerA, and \fear more challenging, their performance remains stable and never drops to the critical levels seen in LLM outputs. In contrast, LLMs significantly underperform on these emotions, with pronounced precision-recall imbalances and frequent misclassifications. This suggests that LLMs struggle with subtle or ambiguous emotions that humans, despite some variation, handle more reliably.

Additionally, when interpreting correctness, it is important to note that our LLM annotation process was designed to rely on independent LLM instances, without mechanisms for discussion or disagreement resolution, unlike human annotators in our study. As a result, 39 sentences (3.5\%) in LLM agreement ($LLM_A$) were left without an assigned label, since inter-LLM discussions, akin to human deliberation, were beyond the scope of this study. Similarly, before discussion, human annotations led to 34 sentences (3.1\%) without an assigned label, which were later resolved through deliberation. Integrating discussion mechanisms into LLM annotation workflows could enhance performance~\cite{kim-etal-2024-meganno}, particularly for ambiguous emotions where disagreement is more prevalent.

\subsubsection{Cost-efficiency Analysis} 
%Table~\ref{tab:cost-efficiency} reports the cost-efficiency comparison between human and LLM-based annotators\footnote{Assuming one human annotator participating in all iterations for better comparability with LLM-based annotations.}, using average values per annotator for better comparability. 
%% ALTERNATIVA
Table~\ref{tab:cost-efficiency} compares the cost-efficiency of human and LLM-based annotators\footnote{Extended results per annotator are included in the replication package.}. For better comparability, the table reports average time and cost per annotator. For human annotators ($Ann_i$), we compute the average across the three annotators participating in each iteration. For LLM annotators ($GPT_i$, $Mistral_i$, $Gemini_i$), we take the average across the three annotation runs. %For better comparability, the table reports average values per annotator and assumes the cost of a human annotator participating in all iterations.
On average, a human annotator takes more than $7 \times$ longer to annotate the whole dataset ($\approx 7$ hours) than the slowest LLM-based annotator\footnote{LLM annotators integrate annotation guidelines as system prompts during each iteration; therefore, guideline processing time is not reported.}, Gemini ($\approx 1$ hour). Cost disparity is even more pronounced -- on average, one human annotator is $233\times$ more expensive\footnote{Based on the average hourly cost of the five annotators.} than the most costly LLM annotator, GPT-4o. These findings highlight the scalability of LLM-based annotation, offering a %faster and significantly more 
cost-effective alternative to manual annotation, particularly for iterative and large-scale annotation tasks.

\begin{table}[t]
\caption{Cost-efficiency analysis (average per annotator)}
\label{tab:cost-efficiency}
\centering{
%\resizebox{\linewidth}{!}{%
\begin{tabular}{@{}llrrr@{}}
\toprule
\textbf{Metric} & \textbf{Annotator} & \textbf{Guidelines} & \textbf{Iteration} \textit{(n=100)} & \textbf{All} \textit{ (n=1112)} \\ 
\midrule
Time & Ann\textsubscript{i} & 11.4$'$  & 38.4$'$ & 427.0$'$ \\
 & GPT\textsubscript{i} & - & 2.9$'$& 31.9$'$ \\
  & Mistral\textsubscript{i} & - & 2.2$'$ & 24.7$'$ \\
   & Gemini\textsubscript{i} & - & 5.2$'$ & 57.9$'$ \\\midrule
Cost & Ann\textsubscript{i} & 7.17 € & 24.14 € & 268.4 € \\
 &  GPT\textsubscript{i} & 1.34 € & 0.10 € & 1.15 € \\
  &  Mistral\textsubscript{i} & 1.30 € & 0.07 € & 0.83 € \\
   &  Gemini\textsubscript{i} & 0.06 € & $<$0.01 € & 0.05 € \\ 
\bottomrule
\end{tabular}
}
\end{table}

\section{Discussion}
\label{sec:discussion}

\subsection{Research Questions}

\subsubsection{Which taxonomy of emotions is most suitable for annotating mobile app reviews? (RQ\textsubscript{1})}

We establish Plutchik's emotion taxonomy as the most effective framework for annotating mobile app reviews due to its structured categorization and relevance to key app-related emotions. %Our selection of Plutchik’s emotion taxonomy for annotating mobile app reviews is motivated by findings from the literature review 
Our selection is grounded by a systematic literature review (see Section~\ref{sec:res-literature-review}), focusing on emotion popularity, frequency, and relevance to our domain, particularly for emotions such as \anticipationA, which helps identify feature requests; \trustA, which aligns with user goals and experiences; and \disgustA, which signals critical issues leading to user rejection of an app or feature. 
While our study is based on this taxonomy, the proposed framework is fully adaptable. Researchers can refine annotation guidelines (RQ\textsubscript{2}) and modify input/output formats for LLM-based annotation (RQ\textsubscript{4}). 
Additionally, while our focus is on emotions in app reviews, the literature review was conducted within the broader scope of software analysis. While our guidelines and taxonomy adaptations are tailored to the mobile app context (RQ\textsubscript{2}), the same approach could be extended to other software-related user feedback domains, such as software product reviews or issue tracking systems.

\subsubsection{How can the selected taxonomy be effectively adapted to the specific context of app reviews? (RQ\textsubscript{2})}

Our adaptation of Plutchik's taxonomy addresses the unique characteristics of mobile app reviews by combining structured guidelines with practical examples (Table~\ref{tab:emotion-guidelines}, Section~\ref{sec:guidelines-dataset}). 
%Table~\ref{tab:emotion-guidelines} in Section~\ref{sec:guidelines-dataset} presents a synthesized adaptation of Plutchik's taxonomy, detailing key sentiments, relevant contents, and examples from mobile app reviews. 
The dataset illustrates how these guidelines can be applied effectively,  
%Our annotation guidelines offer an extended analysis of each emotion, 
supporting both replication of our study and reuse of our guidelines and dataset in future research. 

Developing clear, practical and unambiguous annotation criteria proved essential. To enhance practical usability, we iteratively refined our guidelines based on feedback from human discussions (RQ\textsubscript{3}), incorporating disambiguations, additional examples, and specific criteria to distinguish overlapping emotions. For instance, we established clear distinctions between \sadness and \surprise (disappointment vs. unexpectedness) and \joy and \trust (appraisal vs. user engagement), helping annotators make more consistent decisions. Finally, our findings suggest that merging closely related emotions into broader categories could enhance classification performance while preserving interpretability. %This would provide a practical perspective on the applicability of the taxonomy in real-world scenarios, while maintaining the analytical depth established in this study.

\subsubsection{What challenges arise when annotating app reviews with emotional labels? (RQ\textsubscript{3})}

We identified four key challenges in emotion annotation (see Section~\ref{sec:res-challenges}). While these challenges primarily pertain to human annotation, they are also assessed in the context of LLM-based automated annotation (see discussion on RQ\textsubscript{4}). These challenges serve as groundwork for identifying design strategies to support automated emotion extraction approaches. Mainly, we propose the following: 

\begin{enumerate}
    \item \textbf{Incorporating confidence scores} and \textbf{human-in-the-loop mechanisms} to prioritize high-certainty emotions while flagging low-confidence predictions. This mitigates the difficulty of defining boundaries between contiguous emotions (\textit{Challenge 1}) and helps establish thresholds for subtle emotional intensity by reducing overinterpretation and underreporting (\textit{Challenge 3}).
    
    \item \textbf{Leveraging attention-based architectures} (e.g., BERT~\cite{Devlin2019}) with \textbf{explainability techniques} (e.g., SHAP~\cite{Lundberg2017}) to improve transparency and traceability of mixed or conflicting emotions. This ensures better interpretability of overlapping emotional signals within a single sentence (\textit{Challenge 2}).
    
    \item \textbf{Employing context-enhanced input pipelines} that integrate full reviews and external metadata (e.g., app name, category, or user rating) to improve emotion prediction accuracy. This helps address the lack of context and linguistic ambiguities that hinder sentence-level emotion extraction (\textit{Challenge 4}).
    
    \item \textbf{Using multi-label classifiers or ensembles of binary classifiers} to better capture complex emotional expressions. This ensures that non-mutually exclusive emotions can be effectively modeled, particularly when multiple emotions co-occur in the same sentence (\textit{Challenge 2}).
\end{enumerate}

%\textit{(i)} employing multi-label classifiers or ensembles of binary classifiers to improve emotion-level performance; \textit{(ii)} leveraging advanced attention-based models, such as Transformer architectures and LLMs; \textit{(iii)} integrating explainability techniques (e.g., SHAP~\cite{Lundberg2017}) to trace emotion labels back to specific sentence components; \textit{(iv)} ; and \textit{(v)} implementing context-enhanced pipelines capable of processing complex input structures beyond isolated sentences. 
%These challenges and design suggestions underscore the need for adaptive rameworks that blend automation with human oversight, improving reliability and interpretability in emotion classification for app reviews.

\subsubsection{How does LLM-based annotation compare to human annotation in emotion classification for app reviews? (RQ\textsubscript{4})}

While LLMs provide a cost-effective alternative, human annotation remains the most reliable reference point, reinforcing their role as ground truth. Although LLMs exhibit internal consistency, they do not perfectly align with human interpretations, needing calibration or fine-tuning for domain-specific tasks. Using LLM annotations as a substitute for human annotation requires validation, as their agreement with human labels is moderate but not exact.

LLM-based agreement and correctness assessment demonstrates that the challenges affecting human annotation (RQ\textsubscript{3}) also apply to LLMs. Poor performance in \anger and \disgust correlates with the difficulty of delineating overlapping negative emotions (\textit{Challenge 1}). Similarly, frequent misclassification of \surprise suggests ambiguity in emotional intensity and context, reflecting human annotators’ struggles in defining arousal thresholds (\textit{Challenge 3}). Furthermore, the instability in low-precision classes highlights the limitations of sentence-level annotation without broader contextual cues, emphasizing the need for context-enhanced pipelines (\textit{Challenge 4}).

Although we evaluate LLM correctness relative to a human-annotated ground truth ($Ann_A$), our study does not assess the performance of LLMs in fully automated emotion extraction within user feedback analysis pipelines. Instead, we focus on the accuracy of LLMs as an alternative to human annotators, examining the cognitive challenges and limitations humans face when constructing foundational resources (e.g., an annotated dataset) for emotion classification. 

Future research can reuse our dataset ($Ann_A$) to support automatic emotion extraction in various LLM-based settings. Encoder-only LLMs, such as BERT, RoBERTa, or XLNet, could be applied in a supervised text classification setting. Alternatively, decoder-only -- generative -- LLMs could be employed via fine-tuning or few-shot prompt engineering using partial subsets of our dataset. Further investigations should explore and compare multiple LLM-based annotation strategies, capitalizing on the ground truth, challenges, and design recommendations derived from this study.

\subsection{Research Goal and Implications}

Our study showcases how fine-grained emotion analysis in mobile app reviews faces three core limitations: (1) the lack of domain-adapted taxonomies, (2) the ambiguity in human interpretation of emotion labels, and (3) the uncertain reliability of automated methods. Our study directly addresses these limitations by operationalizing a widely accepted taxonomy (i.e., Plutchik's) for the mobile app domain, developing annotation guidelines that clarify emotion boundaries with domain-specific examples, and empirically demonstrating where LLMs align or diverge from human annotation. These contributions not only make emotion classification more feasible and interpretable but also expose the practical constraints and design trade-offs in scaling annotation processes—offering grounded insight for future tool development and dataset construction.

As a result, our study unlocks new opportunities in traditional RE tasks within the context of opinion mining from app stores. Our contributions (i.e., annotation guidelines, dataset, automated annotation pipelines) serve as groundwork for building (semi-)automatic pipelines for more precise categorization of user feedback, enabling a diversity of use cases that can be explored by leveraging the artifacts resulting from this research—such as feature-emotion pair sentiment analysis (UC\textsubscript{1}), emotionally-oriented user segmentation for release planning (UC\textsubscript{2}), and fine-grained issue prioritization (UC\textsubscript{3}). The analysis of the generated dataset (RQ\textsubscript{2}) and the challenges identified during its generation (RQ\textsubscript{3}) demonstrate how emotional labels such as \anticipationA, \trustA, \surpriseA, or \fear offer extended knowledge dimensions compared to traditional polarity-based approaches. Furthermore, emotion analysis not only conveys the insights captured by polarity (i.e., positive, negative) and review type (i.e., bug report, feature request, praise~\cite{Maalej2015}) classification pipelines, but also expands on additional emotional dimensions (e.g., \surpriseA, \fearA, \trustA) that might otherwise be neglected.

\subsection{Threats to Validity}

Concerning threats to validity~\cite{Wohlin2012}, annotators relied on the full review to resolve linguistic ambiguities, which may have introduced unintended dependencies on contextual information. While this approach improved consistency, it also introduced the risk that certain emotional cues may have been interpreted differently when considering broader review contexts. 
Additionally, personal bias during human annotation posed a threat, particularly when annotators encountered ambiguous or overlapping emotional expressions. While involving five annotators and the generation, discussion, and refinement of the guidelines reduced this risk, differences in annotator interpretations may still have led to slight variations in the annotated dataset. 
Furthermore, annotation challenges such as distinguishing subtle variations in intensity and resolving conflicting emotions (e.g., \joy vs. \trustA, \sadness vs. \disgustA) may have affected annotation reliability. To address this, we specifically addressed these challenges through dedicated discussions and mitigation actions to improve the guidelines. Furthermore, we report these challenges to assist further research in considering the limitations of our dataset properly.% (RQ\textsubscript{3}).

The selection of Plutchik’s taxonomy might not fully capture all possible emotional states present in app reviews. We mitigated this risk by conducting a thorough literature review before selecting the taxonomy. In addition, as our research design remains agnostic to a specific taxonomy, our replication package enables adaptation to alternative emotion taxonomies and annotation schemes. Another limitation from the annotation process is the assessment of emotional intensity, as subtle arousal variations remain difficult to measure consistently.%, highlighting the need for more precise annotation methods in future work.

Finally, our dataset may not generalize to all app review domains. Although reviews span diverse categories and features, further validation is needed to assess applicability across other domains, languages, and cultural contexts. Linguistic differences and user demographics may affect how emotions are conveyed in app reviews, suggesting the need for cross-domain evaluation. Regarding LLM selection, our results may not fully generalize to other architectures or newer models. Exploring alternatives like DeepSeek %\footnote{\href{https://www.deepseek.com/}{https://www.deepseek.com/}} 
(with restricted API access during this research) or OpenAI reasoning models (e.g., o1), now available via API, could yield different outcomes, potentially improving agreement or introducing new annotation biases.
\section{Related Work}
\label{sec:related-work}

\subsection{Emotion Annotation of App Reviews}

The use of multi-class, fine-grained emotion taxonomies in app reviews remains limited, though some related work exists. Riccosan published an Indonesian dataset of app reviews annotated with Parrott's taxonomy~\cite{Riccosan2023}. While relatively large (20K reviews), it is not available in English. Moreover, their annotation relied on two annotators, reporting a Cohen’s Kappa of $0.61$. They lack details on disagreement resolution and insights into how specific emotions were adapted to the app review domain. Moreover, emotions like \anticipation and \trustA, relevant to app reviews, are not considered.

Several studies have explored methods for inferring emotions from app reviews using proprietary datasets. Malgaonkar et al. developed a tool integrating a WordNet-based lexicon method to identify Ekman’s emotions from a dataset of 53K reviews~\cite{Malgaonkar2019}. Similarly, Keertipati et al. applied a lexicon-based method using the LIWC dictionary for three negative emotions to analyse their correlation with app features~\cite{Keertipati2016}. Singh et al. manually annotated 2K mobile learning app reviews using also a lexicon-based approach aligned with Plutchik’s taxonomy~\cite{Singh2022101929}, linking emotions to review descriptors like ratings, technical quality and usefulness. Beyond lexicon-based methods, Savarimuthu et al. employed IBM Watson’s Tone Analyzer %\footnote{\href{https://cloud.ibm.com/docs/natural-language-understanding}{https://cloud.ibm.com/docs/natural-language-understanding}} 
to extract emotions as descriptors for assessing data waste in mobile app reviews~\cite{Savarimuthu2023}. Lastly, Cabellos et al.~\cite{Cabellos2022} manually analysed video game reviews using Liew \& Turtle’s taxonomy to align emotions with moral aspects. However, these datasets are unavailable, lack emotion annotations, and do not evaluate extraction methods empirically. This underscores the relevance of emotion analysis but highlights the scarcity of annotated datasets.

\subsection{LLM-based Annotation}

The potential of LLMs for human-like reasoning tasks, combined with the need for large domain-specific datasets to reduce hallucinations and errors, has driven research into their use as data annotators~\cite{Hou2024}. Heseltine et al. analysed the performance of multiple annotation runs using OpenAI's GPT-4 for political text annotation~\cite{Heseltine2024}. Their findings suggest that while LLM-assisted tagging achieves high accuracy for simple tasks in cost-efficient settings, it struggles with complex and subjective analyses, such as sentiment annotation, where interpretation often varies between annotators~\cite{Zhang2025}. Similarly, Sayeed et al. evaluated Gemini for text classification in materials science, reaching comparable conclusions~\cite{Sayeed2024}. Research has further explored LLM annotation across various fields, including mathematics~\cite{Shan2024}, finance~\cite{Aguda2024}, and linguistics~\cite{Yu2024}. 

To address these limitations, Kim et al. proposed MEGAnno+~\cite{kim-etal-2024-meganno}, a \textit{human-LLM} collaborative framework designed to enhance the reliability and robustness of LLM-generated labels. Their approach integrates a human-in-the-loop mechanism to verify LLM annotations, concluding that fully autonomous annotation remains prone to errors, requiring human oversight for reliability. Similar studies investigate additional dimensions, such as the explainability~\cite{Wang2024} and cost-effectiveness~\cite{Rouzegar2024} of human-LLM collaboration in annotation tasks. While several domain-specific studies have been conducted, further research is needed to assess the reliability of these agents and explore improvements through alternative annotation mechanisms. To this end, and in line with our findings, hybrid approaches that combine expert validation with automated annotations may provide a balanced solution for generating datasets to support supervised extraction methods.
\section{Conclusions}

Our study addresses key challenges in fine-grained emotion extraction from mobile app reviews, offering several contributions: (i) clear emotion annotation guidelines tailored to mobile app reviews, (ii) an annotated dataset that adapts Plutchik’s taxonomy to this domain, (iii) an explicit identification of key challenges in this task, and (iv) a comparison between human and LLM-based annotations to assess the feasibility of scalable, automated annotation using LLMs. Together, these insights and data-based resources are readily available to support future research and advance progress in this field. %While LLMs reduce manual effort, their agreement with human judgment remains moderate, requiring further improvements or human supervision for reliable automated annotation.

%Our research aims to advance fine-grained emotion extraction from mobile app reviews by providing a structured dataset and annotation guidelines for fine-grained emotion extraction from mobile app reviews, adapting Plutchik’s taxonomy to this domain. By comparing human and LLM-based annotation, we highlight key challenges in emotion classification, including ambiguity, data imbalance, and context dependence. While LLMs offer a scalable alternative to manual annotation, their agreement with human judgment remains moderate, requiring further improvements or human supervision for automated annotation.

Future work will focus on automating emotion extraction using LLMs and mitigating data imbalance. We plan to explore fine-tuned encoder-based models (e.g., BERT, RoBERTa) and evaluate generative LLMs with few-shot learning and retrieval-augmented generation. Additionally, we will investigate data augmentation techniques such as synthetic data generation and paraphrasing to improve classifier robustness. Overall, our work sets the stage for improving automated emotion extraction through enhanced models and data strategies.
\section*{Data Availability Statement}
\label{sec:das}

Full datasets, annotations, agreement results, and code for LLM-based annotation are available at \href{https://doi.org/10.6084/m9.figshare.28548638}{https://doi.org/10.6084/m9.figshare.28548638}. % and at \href{https://github.com/nlp4se/review-emotion-analysis}{https://github.com/nlp4se/review-emotion-analysis}.
\section*{Acknowledgments}

%This project has received funding from the European Union's Horizon Europe research and innovation programme under grant agreement Nº 101189745 - HIVEMIND.

This work has been supported by funding from the HIVEMIND project – Horizon Europe call HORIZON-CL4-2024-DIGITAL-EMERGING-01 under Grant Agreement Number 101189745. This paper has been funded by the Spanish Ministerio de Ciencia e Innovación under project/funding scheme PID2020-117191RB-I00 / AEI/10.13039/501100011033.

%This project has received funding from the European Union's Horizon Europe research and innovation programme under grant agreement Nº 101189745 - HIVEMIND.
% \section*{AI Usage Statement}
% The authors used OpenAI's GPT-4o model for language polishing and grammar refinement. All conceptualization, research, analysis, and conclusions are solely their own.

\clearpage
\bibliographystyle{IEEEtran}
\bibliography{ref}

\end{document}